\documentstyle[prl,aps,floats,twocolumn,tighten]{revtex}

\def\lsim{\mathrel{\rlap{\lower4pt\hbox{\hskip1pt$\sim$}}
    \raise1pt\hbox{$<$}}}         
\def\gsim{\mathrel{\rlap{\lower4pt\hbox{\hskip1pt$\sim$}}
    \raise1pt\hbox{$>$}}}         

\newcommand{\vecx}{{\mathbf x}}

\long\def\comment#1{}

\begin{document}
\draft
\twocolumn[\hsize\textwidth\columnwidth\hsize\csname @twocolumnfalse\endcsname
\title{Spintessence!  New Models for Dark Matter and Dark Energy}
\author{Latham A. Boyle$^1$, Robert R. Caldwell$^2$, and Marc Kamionkowski$^1$}
\address{$^1$Mail Code 130-33, California Institute of Technology,
Pasadena, CA 91125}
\address{$^2$Department of Physics \& Astronomy, Dartmouth College, Hanover,
NH 03755}
\date{May 2001}
\maketitle

\begin{abstract}

We investigate a class of models for dark matter and/or negative-pressure,
dynamical dark energy consisting of ``spintessence,'' a complex scalar field
$\phi$ spinning in a $U(1)$-symmetric potential $V(\phi)=V(|\phi|)$. As the
Universe expands, the field spirals slowly toward the origin. The internal
angular momentum plays an important role in the cosmic evolution and
fluctuation dynamics. We outline the constraints on a cosmic spintessence
field, describing the properties of the potential necessary to sustain a viable
dark energy model, making connections with quintessence and self-interacting
and fuzzy cold dark matter. Possible implications for the coincidence problem,
baryogenesis, and cosmological birefringence, and generalizations of
spintessence to models with higher global symmetry and models in which the
symmetry is not exact are also discussed.

\end{abstract}

]

\narrowtext

\noindent{\bf Introduction.} 
Supernova evidence \cite{supernovasearches} for an accelerating Universe has
been dramatically bolstered by the discrepancy between the total cosmological
density $\Omega_{\rm tot}\simeq1$ indicated by the cosmic microwave background
(CMB) \cite{CMBold} and dynamical measurements of the nonrelativistic-matter
density $\Omega_m\simeq0.3$. New and independent evidence is provided by higher
peaks in the CMB power spectrum that also suggest $\Omega_m\simeq0.3$
\cite{CMBold}, again leaving 70\% of the density of the Universe unaccounted
for. As momentous as these results are for cosmology, they may be even more
remarkable from the vantage point of fundamental physics since they indicate
the existence of some form of negative-pressure ``dark energy''.

For this dark energy to accelerate the expansion, its equation-of-state
parameter ${w}\equiv p/\rho$ must satisfy ${w}<-1/3$, where $p$ and $\rho$ are
the dark-energy pressure and energy density, respectively. The simplest guess
for this dark energy is the spatially uniform, time-independent cosmological
constant for which ${w}=-1$. Another possibility is quintessence \cite{CDS98},
a cosmic scalar field \cite{quint} that is displaced from the minimum of its
potential. Negative pressure is achieved when the kinetic energy of the rolling
field is less than the potential energy, so that $-1 \le {w} < -1/3$ is
possible.

This negative-pressure dark energy should not be confused with the cold dark
matter that has long been known to be required to support flat galactic
rotation curves and to provide the majority of the matter in galaxy clusters. 
Leading candidates for this dark matter include collisionless particles such as
supersymmetric particles \cite{JunKamGre96} and the axion \cite{axions}. 
However, numerical simulations of structure formation with collisionless dark
matter seem to indicate more galactic substructure than is observed
\cite{dwarfs}, a discrepancy that has led some to postulate that the dark
matter might possess a self-interaction \cite{SpeSte00} or consist of extremely
low-mass particles (``fuzzy'' dark matter) \cite{HuBarGru01}.

In this paper, we consider a new class of models for dark matter and dark
energy. We investigate the behavior of a complex scalar field that is spinning
in a circular orbit in a $U(1)$-symmetric potential $V(\phi)=V(|\phi|)$, a
monotonically increasing function of $|\phi|$. As the Universe expands, the
radius of this orbit, and thus the potential- and kinetic-energy densities
decrease. It is the internal-angular-momentum barrier, not expansion friction,
that prevents the field from falling directly to the minimum of the potential.
Unlike quintessence models, spintessence allows $|\phi|$ to change slowly even
if the time derivative of $\phi$ is large.  As well, the growth of
perturbations in spintessence differs from those in quintessence or cold dark
matter.

Below, we discuss the evolution of spintessence and the growth of
perturbations, working through some simple illustrative examples. We conclude
with some remarks about the viability of spintessence models with global
symmetries other than $U(1)$ or in the presence of broken global symmetry, and
we mention possible links to quintessence, baryogenesis, and other areas of
particle physics and early-Universe cosmology.

\smallskip
\noindent{\bf Spintessence.}
We can decompose a complex scalar field into two real fields:
$\phi(\vecx,t)=\phi_1(\vecx,t) + i \phi_2(\vecx,t) \equiv R(\vecx,t)\exp[i
\Theta(\vecx,t)]$. First suppose that $\phi$ is homogeneous, lives in Minkowski
space, and has a $U(1)$-symmetric potential-energy density $V=V(|\phi|)$ that
is a monotonically increasing function of $|\phi|$. Then its equations of
motion are equivalent to those of a classical particle moving in a
two-dimensional central potential $V(R)$. The simplest non-trivial solutions
are those in which the field moves in a circular orbit, $\phi(t)=R e^{i \omega
t}$, with $R$ and $\omega$ constants that satisfy $R\omega^2 = V'(R)$ so the
centripetal acceleration balances the radial force.

In an expanding Universe, conservation of the global-charge current means
$\dot\Theta = Q/a^3 R^2$ where $Q$ is a constant associated with the total
charge, and $a(t)$ is the cosmological scale factor. With regards to the field
dynamics, the charge introduces a secular driving-term into the
equation-of-motion for $R$,
\begin{equation}
\ddot R + 3 H \dot R + V'(R) = {Q^2 \over a^6 R^3} 
\end{equation}
where $H=\dot a/a$.  If the spin frequency is high, $\dot\Theta\gg H$,  we may
expect the rotation to dominate, supporting the field against radial infall. In
this rapidly-spinning approximation, the time evolution of $R$ is then
determined from $V'(R) = Q^2 a^{-6} R^{-3}$. From this we find that the
potential must satisfy $(d/dR)[R^3 V'(R)]>0$ if it is to be steep enough to
confine the field to a circular orbit as the Universe expands. For instance,
with a quadratic potential, $R \propto a^{-3/2}$ in a matter-dominated epoch so
that the radial kinetic energy rapidly decays $\dot R^2 \propto a^{-6}$,
leaving energy density and pressure
\begin{equation}
\rho = \frac{1}{2}(\dot R^2 + R^2 \dot\Theta^2) + V, 
\quad
p = \frac{1}{2}(\dot R^2 + R^2 \dot\Theta^2) - V,
\end{equation} 
with $R^2\dot\Theta^2 = 2 V \propto a^{-3}$, and an equation-of-state $w=0$.
For such rapidly spinning fields, the equation-of-state parameter is
\begin{equation}
     w(R) \approx {R V'(R) - 2 V(R) \over R V'(R) + 2 V(R)} .
     \label{eos}
\end{equation}
However, solutions with an arbitrary constant equation-of-state, for which each
term in $\rho,\,p$ above decays as $\propto a^{-3(1+w)}$, are not possible
owing to the conserved charge.   

\smallskip
\noindent{\bf Growth of perturbations.}
We now consider the growth of perturbations in spintessence. While the
perturbations in a spinning field have been considered (for different purposes)
in Refs. \cite{KhlMalZel85,JetSci97} for the special case of quadratic and
quartic potentials, here we generalize their analysis to arbitrary potentials.
We start with the spacetime line element
\begin{equation}
     ds^2 = (1+2 \Phi) dt^2 - (1-2\Phi) a^2(t) d\vecx^2.
\end{equation}
where $\Phi(\vecx,t)$ is the Newtonian potential arising from fluctuations in
the spinning field, $R(t) + \delta R(\vecx,t)$ and  $\Theta(t) + \delta
\Theta(\vecx,t)$, and surrounding matter. The evolution equations  for the
perturbations, $\delta R$ and  $\delta\Theta$, obtained from the linearized
Einstein Equations are 
\begin{eqnarray}
\ddot {\delta R} + 3 {\dot a \over a} \dot{\delta R} &&+ 
\big( V'' - \dot\Theta^2 -{1 \over a^2} \nabla^2 \big) {\delta R}
\cr
&&= 4 \dot R \dot\Phi - 2 \Phi V' 
+ 2 R \dot\Theta \dot{\delta\Theta}, \\
\ddot {\delta\Theta} + 3 {\dot a \over a} \dot{\delta\Theta} && - 
{1 \over a^2} \nabla^2 \delta\Theta \cr
&& = 
4 \dot \Theta \dot\Phi 
-2 {\dot{\delta R} \over R} \dot\Theta + 2 {\dot R \over R}
\left({\delta R \over R} \dot\Theta - \dot{\delta\Theta}\right), \\
\nabla^2\Phi - 3 H \dot\Phi &&- 3 H^2 \Phi \cr
&&= 4 \pi G \big[
\dot R \dot{\delta R} + V' \delta R + R^2 \dot\Theta \dot{\delta \Theta} \cr
&&+ R \dot\Theta^2 \delta R 
-\Phi \big(\dot R^2 + R^2 \dot\Theta^2 \big)\big].
\label{poisson}
\end{eqnarray}
The final line gives the constraint equation to the gravitational potential.
The stability of a real scalar field  depends on the effective mass, $V''$. But
here we see that the stability criteria for the spinning field must differ 
since not only is the effective mass different, $V'' - \dot\Theta^2$, but also 
the $\delta R$ and $\delta\Theta$ equations are coupled.  Before proceeding to
a full-blown relativistic calculation, we can infer essential information about
the behavior of perturbations for the rapidly spinning field with a Newtonian
analysis set in Minkowski spacetime. There, perturbations to the gravitational
potential of the form
\begin{equation}
\Phi(\vecx,t) = \Phi_1  e^{\Omega t + i\vec k \cdot\vec x}
\end{equation}
will be generated through the Poisson equation (\ref{poisson})
by small amplitude perturbations to the amplitude and phase of the
scalar field, 
\begin{equation}
\delta R(\vecx,t) = R_1 e^{\Omega t + i\vec k\cdot\vec x}, \quad
\delta\Theta(\vecx,t) = \Theta_1  e^{\Omega t + i\vec k\cdot\vec x}.
\end{equation}
Leaving out terms that are small for $k^2\gtrsim G\rho$, corresponding to
wavelengths inside the horizon, we obtain the wavenumber $k_J$ at which
$\Omega^2=0$.  For $k < k_J$, where $k_J$ is the Jeans wavenumber 
\begin{equation}
     k_J^2 = {1\over 2} \left[ {V' \over R}-V'' + \sqrt{ \left( {V'
     \over R}  - V'' \right)^2 + 64 \pi G V'^2} \right] 
\label{eq:jeans}
\end{equation}
then $\Omega^2 >0$, and the perturbations grow exponentially. For $k>k_J$ then
$\Omega^2< 0$, and the perturbations oscillate in time. Note that these
instabilities will be effective in an expanding Universe as long as we consider
the following: (1) The physical wavelength associated with a given comoving
wavelength changes with time.  (2) Perturbations on scales larger than the
horizon will be stabilized.  (3) The time dependence of unstable perturbations
inside the horizon will be power law rather than exponential.

If the spintessence field is to supply a dark matter component, then the
existence of a gravitational instability, whether exponential or power-law, is
welcome. As a dark energy candidate, however, we require stability against the
growth of perturbations on scales at least as large as clusters. As a complex
field with a conserved charge, $Q$, the spintessence field is susceptible to
the formation of Q-balls --- a nontopological soliton \cite{QBallCrit}. 
In order to avoid the formation of Q-balls from the spinning field, however, it
is necessary that the field does not have an instability that drives $R\to
R_{qb}$, the non-zero value of the field amplitude at which the quantity
$V(R)/R^2$ has a minimum. Checking with (\ref{eos}), we note when $w=0$, and 
provided $V''(R_{qb})>0$, the conditions are ripe for the formation of
Q-balls.  This means that a rapidly spinning field cannot safely pass through a
$w=0$ phase. This consideration places a further constraint on the behavior of
a viable spintessence model of dark energy. Note that if the spinning field
does not entirely decay into Q-balls, there is the interesting possibility that
both dark energy and dark matter might consist of the spinning field, which
would provide an interesting possibility for solving the coincidence  problem.

We now consider specific examples of spintessence potentials.

\smallskip\noindent{\bf Power-Law:} A rapidly spinning field in a potential
$V(R)=V_0 (R/R_0)^n$ has a constant equation-of-state parameter ${w}=(n-2)/
(n+2)$. The quadratic $U(1)$ potential leads to a matter density that decays as
cold dark matter, as the dynamics of $\phi_1$ and $\phi_2$ are those for two
decoupled harmonic oscillators.  Since $V'/R-V''=0$  we find $k_J^2 \simeq 4
\sqrt{\pi G} V'$, and the instability is driven by gravity.  Perturbations on
smaller scales are stabilized by scalar-field dynamics.   Such a field is
unstable to the formation of Q-balls, which may provide for an interesting dark
matter component. On the other hand, potentials with $n>2$ are stable against
Q-ball formation. The Jeans scale is $k_J^2 \simeq 16 \pi G R V'/(n-2)$,
provided $GR^2 \ll 1$. However, $w>0$ so that these are less interesting from
the dark matter / dark energy perspective. 

A dark energy component with a $n<1$ power-law potential is plagued by an
instability which leads to the formation of Q-balls. Since a dominant component
necessarily has $V \sim (m_{pl} H)^2$, we find the Jeans wavenumber is $k_J
\sim H m_{pl}/R$. Although $R$ might start out with an amplitude comparable to
the Planck mass,  since $R$ necessarily decays as the Universe expands, it is
inevitable that the wavenumber will eventually be well within the Hubble
horizon. The behavior $d{k_J}/dR <0$ as $R$ decays disqualifies a wide class of
potentials as dark energy.

\smallskip \noindent{\bf Self-interacting and fuzzy cold dark matter.}  Suppose
$V(R)=\frac{1}{2}m^2 R^2 + \frac{\lambda}{4} R^4$.  If $\lambda>0$, then ${w}=1/3$ at early
times when $\lambda R^4/4 \gg m^2 R^2/2$, but approaches ${w}=0$ at later
times.  For $\lambda<0$  we consider only values of the scalar field
$R<m/\sqrt{-\lambda}$, and in this case, there is a negative pressure that
approaches ${w}=0$ at late times as the quartic term  becomes small.  If the
quartic term is small, then this describes a gas of cold massive particles that
self-interact via a repulsive ($\lambda>0$) or attractive ($\lambda<0$)
potential.  After collapse and virialization of halos, either type of
interaction would give  rise to a plausible self-interacting dark-matter
candidate.  The  homogeneous and perturbation analysis above can be used to
determine how the mean density and perturbations to this type of  dark matter
would evolve with time.

Now consider the fuzzy cold dark matter of Ref. \cite{HuBarGru01}.  They
suppose that halo dark matter consists of a quadratic potential of mass $m$. 
They adopt a value $m\sim10^{-22}$ eV to smooth galactic halos and since this
dark matter must contribute a density $\rho \sim(10^{-3} \, {\rm eV})^4 \sim
m^2 R^2$, they must have $R \sim 10^{16}\, {\rm eV}$.  This gives rise to a
wavenumber $k_J \sim 10^{-28}$ eV.  More generally, however, there should be a
nonzero quartic term in the potential, but if the dark matter is to be cold,
the quartic term must be small compared with the quadratic term. This leads to
a constraint $|\lambda| \lesssim 10^{-76}$.  As small as this is, the condition
of validity $[(V'/R)-V'']^2 \ll  64 \pi G V'^2$ for their estimate of the Jeans
scale is even more restrictive; it leads to $|\lambda| \lesssim 10^{-87}$.  
Thus, if $10^{-87} \lesssim |\lambda| \lesssim 10^{-76}$, then the Jeans
wavenumber is $10^{-28} \lesssim k_J \lesssim 10^{-22}$ eV for $\lambda<0$, or
$10^{-34} \lesssim k_J \lesssim 10^{-28}$ eV for  $\lambda>0$.  Thus, the
inclusion of a small nonzero quartic interactions can spread the Jeans length
over 11 orders of magnitude.

\smallskip\noindent{\bf Dark Energy:} A spintessence field must meet a number
of constraints in order to be considered as a viable dark energy candidate. 
All together, these conditions may be summarized as: $0 < R V' < V$ to ensure
the existence of circular orbits (lower bound) and equation of state $w<-1/3$
(upper bound); $-\frac{1}{3} R^2 V'' < R V' < R^2 V''$ requiring steep orbits
(lower) and stable perturbations in the absence of gravity (upper). In the
presence of gravity, of course, we use  $k_J \lesssim H$ with equation
(\ref{eq:jeans}) in order to assess stability. Recall that $V(R)$ need not
satisfy these conditions for all $R$, just in the range $R_{max} > R >
R_{min}$, the values at which field evolution begins at early times, and the
value today. Potentials which satisfy the above criteria include $V(R) = M^2
R^2(A + (R/B)^{-r}){\rm e}^{-1/R^2}$ with $1 < r < 2$, or $V(R) =  (M^2 R^2 -
A){\rm e}^{- B R^2} + A$ both with $A,\, B > 0$, as suggested by Kasuya
\cite{Kasuya}.   Along the same lines, a potential of the form $V = M^4
\exp[m^2/(R_{max}^2 - R^2)]$ can give rise to a stable, dark energy component.
In the regime $R < min(R_{max},\, R_{qb})$ where $R_{qb} = - m/2 +
\sqrt{R_{max}^2 + m^2/4}$, the field can evolve for a long time with $w<0$,
before it is necessary to patch on a different functional form for $V$ at some
small value of the field amplitude, say $R_{min}$, to ensure $V(0)=0$. A
stability analysis reveals that the quantity $V'/R - V''$ is negative, unlike
the power-law potential, which immediately tells us from equation
(\ref{eq:jeans}) that gravity will play the dominant role in determining the
Jeans wavenumber. Plugging in our potential, we find $k_J \sim M^2
\sqrt{R}/(m_{pl} R_{max})$. Since a dominant component has $V \sim (m_{pl}
H)^2$, then the Jeans wavenumber reduces to $k_J \sim H \sqrt{m_{pl}
R}/R_{max}$ which is substantially outside the Hubble horizon for $R \ll
R_{max}$. The perturbations are stable. Lastly, there is a novel twist to a
dark energy scenario based on the above potentials. When the field passes to $R
< R_{min}$, the accelerated expansion ends. If the potential in this regime
possesses a minimum in $V/R^2$, then the dark energy field will ultimately
decay into Q-balls.

\smallskip \noindent{\bf Discussion.}  We have considered a class of models for
dark energy and dark matter that consists of a complex scalar field spinning in
a $U(1)$ potential.  Specification of $V(R)$ determines the scaling of the
equation-of-state and density as a function of redshift, and it also determines
how density perturbations grow. These solutions are valid  if the spin
frequency is $\dot\Theta \gg H$.   If the spin period is small,  $\dot\Theta
\lesssim H$, then the field will act like quintessence and will undergo
friction-dominated slow rolling toward the minimum of the potential.  Thus,
depending on the potential, a model may begin as quintessence and wind up like
spintessence, or {\it vice versa}. Fluctuations in spintessence will differ
from a real scalar field, nevertheless, due to the greater number of excitable
degrees of freedom.  Spintessence could conceivably be used to drive inflation
\cite{JetSci97}, although it is difficult to see how the large global charge
density, or alternatively high spin frequency, could be maintained during the
many $e$-folds of expansion required for inflation. Perhaps a greater
difficulty is how to set up a homogeneous, spinning field at the end of
inflation. 

If the dark energy is due to spintessence, then the Universe is in an unstable
state that breaks $T$ and $C$ invariance.  This suggests  interesting
connections between the dark-energy problem  and other questions in cosmology
and particle physics.  For example, if the global charge of the spintessence
field is identified with baryon number, then spintessence may be the vacuum
that hides the antibaryons in a baryon-symmetric baryogenesis model
\cite{DodWid90}. Alternatively, spintessence could conceivably drive
baryogenesis in an Affleck-Dine or spontaneous-baryogenesis model
\cite{AffDin85}.  If the field is coupled to the pseudoscalar of
electromagnetism, it could give rise to $P$- and $T$-violating rotations of
polarization of cosmological sources \cite{Car98}.  

Spintessence should work for higher global symmetries [e.g., $O(N)$ with
$N>2$], as orbits are still confined to a surface in the internal space in such
models. Although heuristic arguments suggest that quantum gravity should
violate global symmetries at least to some degree \cite{KamMar92}, the basic
idea of spintessence should still work. As a simple example, suppose that
$V(\phi_1,\phi_2)= c_1 \phi_1^n + c_2 \phi_2^n$ with $c_1 \neq c_2$.  Although
orbits in this potential are not circular and there is no conserved internal
angular momentum, the virial theorem guarantees that when averaged over an
orbit, the potential-energy density $T$ and kinetic-energy density $V$ will
still be related by $T=(n/2)V$. Thus, as long as the dynamical time for the
potential is small compared with the expansion time, the equation-of-state
should still behave like that for spintessence.

In summary, spintessence is the simplest example of a cosmological field with a
non-trivial internal symmetry group. We have outlined the constraints which
must be satisfied to obtain a viable cosmological model, making connections
with quintessence and varieties of fuzzy- and self-interacting dark matter. We
have shown that adding the internal symmetry gives rise to a rich collection of
new phenomena: different clustering properties, an instability to Q-balls, and
a new way to drive the acceleration, via the angular momentum barrier as
opposed to Hubble friction. If Q-balls or similar objects are an inevitable
by-product of a cosmological field with a non-trivial internal symmetry group,
then it would seem that the viability of spintessence relies on the
compatibility of Q-balls with cosmology.

\smallskip LB was supported by an NSF Graduate Fellowship.  This work was
supported at Caltech by NSF AST-0096023, NASA NAG5-8506, and DoE
DE-FG03-92-ER40701 and DE-FG03-88-ER40397. The work at Dartmouth was supported
by NSF-PHY-0099543.

\smallskip Note: During the preparation of this paper, several other papers
\cite{Kasuya,GuHwa01,Arbey:2001qi,Li:2001xa,Chiueh:2001ri} appeared that also
consider a spinning complex scalar field with respect to dark matter and dark
energy.

\end{document}